\documentclass[aps,prl,twocolumn,superscriptaddress,nofootinbib]{revtex4-2}

\usepackage{graphicx}
\usepackage{dcolumn}
\usepackage{bm}
\usepackage{amsmath}
\usepackage{amsfonts}
\usepackage{amssymb}
\usepackage{physics}
\usepackage{float}
\usepackage[citecolor=blue,linkcolor=blue,urlcolor=blue]{hyperref}
\usepackage[ocgcolorlinks]{ocgx2}
\usepackage{placeins}
\usepackage{multirow}
\usepackage{dcolumn}

\newcommand{\mev}{\,\mathrm{MeV}}

\newcommand{\fmmt}{\,\mathrm{fm}^{-3}}

\newcommand{\prlsection}[1]{\emph{#1.}--}

\begin{document}

\title{Nuclear equation of state for arbitrary proton fraction and temperature\\
based on chiral effective field theory and a Gaussian process emulator}

\newcommand{\tud}{\affiliation{Technische Universit\"at Darmstadt, Department of Physics, 64289 Darmstadt, Germany}}
\newcommand{\emmi}{\affiliation{ExtreMe Matter Institute EMMI, GSI Helmholtzzentrum f\"ur Schwerionenforschung GmbH, 64291 Darmstadt, Germany}}
\newcommand{\mpi}{\affiliation{Max-Planck-Institut f\"ur Kernphysik, Saupfercheckweg 1, 69117 Heidelberg, Germany}}

\author{J.~Keller}
\email{j.keller@theorie.ikp.physik.tu-darmstadt.de}
\tud
\emmi
\author{K.~Hebeler}
\email{kai.hebeler@physik.tu-darmstadt.de}
\tud
\emmi
\mpi
\author{A.~Schwenk}
\email{schwenk@physik.tu-darmstadt.de}
\tud
\emmi
\mpi

\begin{abstract}
  We calculate the equation of state of asymmetric nuclear matter at
  finite temperature based on chiral effective field theory
  interactions to next-to-next-to-next-to-leading order. Our results
  assess the theoretical uncertainties from the many-body calculation
  and the chiral expansion.  Using a Gaussian process emulator for the
  free energy, we derive the thermodynamic properties of matter
  through consistent derivatives and use the Gaussian process to
  access arbitrary proton fraction and temperature. This enables a
  first nonparametric calculation of the equation of state in beta 
  equilibrium, and of the speed of sound and the symmetry energy at finite 
  temperature. Moreover, our results show that the
  thermal part of the pressure decreases with increasing densities.
\end{abstract}

\maketitle

\prlsection{Introduction}
The nuclear equation of state (EOS) plays a central role for the physics
of nuclei and dense matter in neutron stars, supernovae, and 
mergers~\cite{Latt12esymm,Hebeler2015,Lynn2019,Drischler2021ARNPS,Lattimer2021}.
While first principle calculations for a wide range of densities, electron
fractions, and temperatures are desirable, practical calculations are 
limited by uncertainties of the interactions in dense matter and the 
included degrees of freedom. At nuclear densities $n \sim n_0$ (with 
saturation density $n_0=0.16\fmmt$), neutrons and protons are the relevant 
degrees of freedom and chiral effective field theory (EFT) provides a 
systematic expansion of the strong interactions among 
nucleons~\cite{Epelbaum_et_al_2009,Machleidt_Entem_2011,Hammer2020}. 
This has enabled first principle studies of the EOS using various many-body approaches and including theoretical 
uncertainty estimates~\cite{HebelerSchwenk2010,Hebeler_et_al2011,Tews13N3LO,Holt13PPNP,Carb13nm,Hage14ccnm,Coraggio_et_al_2014,PhysRevC.89.064009,PhysRevC.92.015801,Lynn16QMC3N,Dris16asym,Ekst17deltasat,Carbone_et_al_2018,Drischler2019,CarboneSchwenk2019,Lu2020,Drischler2020PRL,Keller2021},
where most calculations have focused on pure neutron matter or symmetric 
nuclear matter. In particular, the combination of chiral EFT results 
for neutron matter and neutron star observations has led to important 
constraints for the EOS in astrophysics and for the properties of neutron stars~\cite{Hebeler2013,Anna18TidalDef,Most18TidalDef,Tews18TidalDef,Lim2018,Capa20NatAst,Essick2020,Dietrich2020,Raaijmakers2021,Essick2021,DrischlerHan2021,Huth2021}.

In this Letter,
we calculate the EOS for arbitrary proton fractions and temperatures based on chiral EFT interactions to high order.
We then construct a Gaussian process emulator of the free energy that 
enables nonparametric evaluations of the EOS and thermodynamic
derivatives for arbitrary nuclear conditions, including beta equilibrium, to
provide direct results for neutron star matter based on chiral EFT.

\begin{figure*}[t]
    \centering
    \includegraphics[clip=,width=\textwidth]{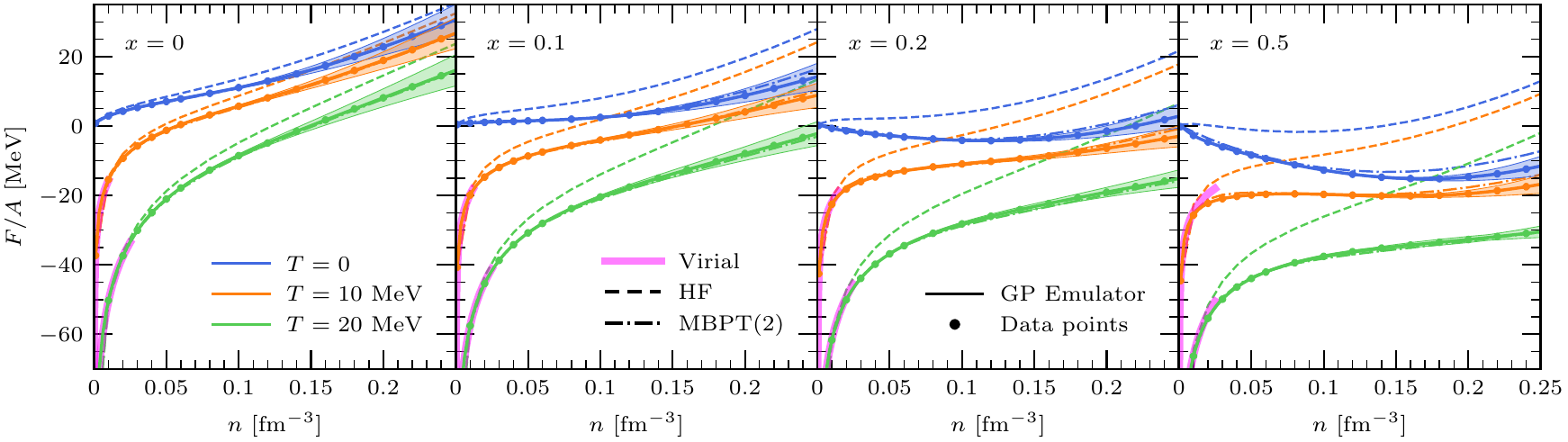}
    \caption{Free energy per particle $F / A$ at N$^3$LO ($\Lambda =
      450\mev$) for different proton fractions $x=0,0.1,0.2$, and $0.5$ 
      (panels from left to right) and for temperatures $T=0,10,$ and 
      $20\mev$ (blue, orange, and green) 
      as a function of density $n$. Our MBPT results are given by 
      the dots, while the constructed GP emulator is shown with solid 
      lines. The bands display theoretical EFT uncertainty
      estimates according to Eq.~\eqref{eq:ekm}.
      To show the MBPT convergence, results at the HF level
      (dashed) and at second order (dot-dashed) are given as well.
      At low densities $n \leqslant 0.025\,\text{fm}^{-3}$,
      we also compare to the virial
      EOS~\cite{HorowitzSchwenk2006a,HorowitzSchwenk2006b} (for 
      $T=20\,\text{MeV}$ this corresponds to a neutron fugacity
      $z_n \leqslant 0.45, 0.39, 0.34, 0.18$ for $x=0, 0.1, 0.2, 0.5$).}
    \label{fig:free-energy}
\end{figure*}

\prlsection{Methods}
Our asymmetric nuclear matter calculations are based on many-body perturbation theory (MBPT) around a self-consistent Hartree-Fock (HF) state. The framework for evaluating MBPT
diagrams using Monte Carlo integration is based on our previous works~\cite{Drischler2019, Keller2021}. We start from the
grand-canonical potential,
\begin{align}
\Omega (\mu_n, \mu_p, T) = -T \ln{\mathrm{Tr}\left(e^{-(H - \mu_n n_n V - \mu_p n_p V) / T}\right)} \, ,
\end{align}
with temperature $T$, volume $V$, and $\mu_{\tau=n, p}$ are the neutron and proton chemical
potentials with corresponding densities $n_\tau$. The Hamiltonian $H = H_0 + V_\mathrm{NN} + V_\mathrm{3N}$ contains a
kinetic term together with nucleon-nucleon (NN) and three-nucleon (3N) interactions constructed from chiral EFT up to next-to-next-to-next-to-leading order (N$^3$LO). The MBPT series at finite $T$ is organized following Refs.~\cite{FetterWalecka,NegeleOrland} with the same choice of reference system as in
Ref.~\cite{Keller2021}. We include all contributions from NN and 3N interactions up to second order, and at third order all interaction vertices that are NN or 3N with one line closing on itself (corresponding to the normal-ordered two-body approximation).
This has been shown to be a very good approximation at $T=0$~\cite{Drischler2019} and for neutron matter at finite $T$~\cite{Keller2021}. For our main results, we employ the NN interactions of Entem, Machleidt, and Nosyk (EMN) with cutoff $\Lambda = 450\,\text{MeV}$~\cite{EMN2017} and 3N interactions fit to nuclear saturation at N$^2$LO ($c_D = 2.25, c_E = 0.07$) and N$^3$LO ($c_D = 0, c_E = -1.32$)~\cite{Drischler2019}.
We include NN partial waves up to total angular momenta $J_{12} \leqslant 12$, and 3N channels up to $J_\text{tot} \leqslant 9/2$ and $J_{12} \leqslant 6$~\cite{Hebeler2021}. These truncations lead to uncertainties that are small compared to the EFT uncertainties for the considered densities.

The free energy density $F/V$ is determined by
\begin{align}
  \frac{F}{V}(n, x, T) = \frac{\Omega}{V}(\mu_n, \mu_p, T) &+ \mu_n n_n(\mu_n, \mu_p, T)\nonumber\\
                                       &+ \mu_p n_p(\mu_n, \mu_p, T) \,,
  \label{eq:free-energy}
\end{align}
where the densities are given by $n_\tau = - \partial_{\mu_\tau}\Omega/V$, the total density is $n = n_n + n_p$, and $x = n_p / n$ is the proton fraction.
To obtain the free energy,
Eq.~\eqref{eq:free-energy}, as a function of density, we invert the
relation between densities and chemical potentials by generalizing the
method from Ref.~\cite{KohnLuttinger1960} to multiple chemical
potentials. In doing so, we formally expand the chemical potentials
around a reference system with the same density and proton fraction as
the interacting system. This re-expansion is necessary to
obtain a perturbation series that is consistent with the
zero-temperature formalism, and effectively deals with the anomalous
diagrams at finite $T$~\cite{KohnLuttinger1960,Wellenhofer:2018qqw, Keller2021}.

As the evaluation of MBPT diagrams involves the computation of
high-dimensional phase-space integrals, 
the computation of the thermodynamic potential for a large number of densities, temperatures and proton fractions is a complex task.
Hence, for the evaluation of the free energy per particle and its derivatives, we construct an emulator for $F(n, x, T)/A$ using three dimensional
Gaussian process (GP) regression~\cite{Rasmussen2005}. Gaussian processes allow us to interpolate the EOS in a
way that does not spoil thermodynamic consistency (e.g., second-order
derivatives commute) and to handle residual
noise from the Monte Carlo integration. We use the Python library of
Ref.~\cite{Chilenski_2015} and employ the squared exponential kernel~\cite{Rasmussen2005} with an overall scale and three length scales as hyperparameters that are determined by maximizing the likelihood.
In constructing the GP, we assume that each diagram has $\Delta E_d = 5\,\text{keV}$ noise from the
Monte Carlo integration and the total noise of every EOS point is
calculated as $\sqrt{\sum_d \Delta E^2_d}$ where the sum is over all
diagrams. The resulting total noise is much smaller than
interaction uncertainties due to the chiral EFT expansion and is not visible in the plots. We treat the Fermi gas (FG) contribution
analytically and emulate the interaction energy per particle
$F_\text{int}/A = F/A - F_\text{FG}/A$. The
GP emulator can be performed in any set of variables. However, replacing $n$ by the Fermi momentum $k_{\text{F}} = (3 \pi^2 n / 2)^{1/3}$ was found to simplify the evaluation of derivatives. Moreover, all input variables are normalized to $[0,1]$ to prevent numerical artifacts in the GP.

\newcolumntype{d}{D{.}{.}{4.4}}
\begin{table}[t]
  \begin{ruledtabular}
    \begin{tabular}{cD{.}{.}{1.2}cdddd}
      \multirow{2}{*}{$T$} & \multirow{2}{*}{$n$} & \multirow{2}{*}{MBPT} & \multicolumn{2}{c}{$x = 0.3$} & \multicolumn{2}{c}{$x=0.5$} \\
       &&& \multicolumn{1}{c}{N$^2$LO} & \multicolumn{1}{c}{N$^3$LO} & \multicolumn{1}{c}{N$^2$LO} & \multicolumn{1}{c}{N$^3$LO} \\
    \hline
    0           & 0.1         & HF           & -0.6           & 0.8            & -3.1           & -1.6           \\
    0           & 0.1         & 2           & -8.7           & -8.4           & -12.4          & -12.1          \\
    0           & 0.1         & 3           & -8.9(5)        & -8.8(2)        & -12.9(4)       & -12.7(1)     \\
    \hline
    0           & 0.16        & HF           & 3.1            & 4.6            & -0.1           & 1.4            \\
    0           & 0.16        & 2           & -8.5           & -8.3           & -13.4          & -13.1          \\
    0           & 0.16        & 3           & -10.1(11)      & -9.9(5)        & -15.5(9)       & -15.1(4)     \\
    \hline
    0           & 0.2         & HF           & 7.9            & 9.1            & 4.5            & 5.5            \\
    0           & 0.2         & 2           & -6.1           & -6.1           & -11.6          & -11.5          \\
    0           & 0.2         & 3           & -8.6(27)       & -8.7(13)       & -14.6(24)      & -14.7(10)     \\
    \hline
    20          & 0.1         & HF           & -23.4          & -23.3          & -26.3          & -26.1          \\
    20          & 0.1         & 2           & -33.6          & -34.1          & -37.7          & -38.1          \\
    20          & 0.1         & 3           & -33.3(9)       & -33.4(5)       & -37.6(8)       & -37.5(4)     \\
    \hline
    20          & 0.16        & HF           & -13.5          & -13.7          & -16.7          & -17.1          \\
    20          & 0.16        & 2           & -28.9          & -29.9          & -34.1          & -35.1          \\
    20          & 0.16        & 3           & -29.8(11)      & -29.0(6)       & -35.5(7)       & -34.3(7)     \\
    \hline
    20          & 0.2         & HF           & -6.4           & -7.2           & -9.7           & -10.8          \\
    20          & 0.2         & 2           & -25.7          & -27.3          & -31.5          & -33.1          \\
    20          & 0.2         & 3           & -27.7(17)      & -26.7(9)       & -34.2(11)      & -32.7(9)     \\
    \end{tabular}
  \end{ruledtabular}
  \caption{MBPT convergence of the free energy per particle $F/A$ in MeV at N$^2$LO
    and N$^3$LO for different proton fractions $x$,
    temperatures $T$ in MeV, and densities $n$ in fm$^{-3}$. The
    EFT uncertainties determined by Eq.~\eqref{eq:ekm}
    are given in parentheses for the third-order MBPT results.}
  \label{tab:mbpt-convergence}
\end{table}

\prlsection{Free energy and GP emulation}
In Fig.~\ref{fig:free-energy} we present results for $F/A$ as a function of density for different proton fractions and temperatures.
We evaluate the MBPT diagrams on the non-uniform grid with values
$n = 0.001, 0.01, 0.02, \dots, 0.05, 0.06, 0.08, \dots,
0.32\,\text{fm}^{-3}$, $x = 0, 0.1, \dots, 0.7$, and
$T = 0, 5, 10, 15, 20, 30\,\text{MeV}$.
The EOS points
are marked with dots in Fig.~\ref{fig:free-energy}, while the results obtained from the GP emulator are shown as solid lines. An excellent agreement is evident. 

At low densities and finite $T$, we compare our results to the model-independent
virial EOS~\cite{HorowitzSchwenk2006a, HorowitzSchwenk2006b} in Fig.~\ref{fig:free-energy}. Since we consider homogeneous matter, we do not include the contributions from alpha particles in the virial EOS (i.e., we compare against Ref.~\cite{HorowitzSchwenk2006b} for $n_\alpha=0$)
For $n \leqslant 0.025\,\text{fm}^{-3}$ and low fugacities, we find excellent agreement 
with our results. For higher densities, the inclusion of higher virial coefficients and
effects due to the effective nucleon mass play an important role. 

\begin{figure*}[t]
    \centering
    \includegraphics[clip=,width=\textwidth]{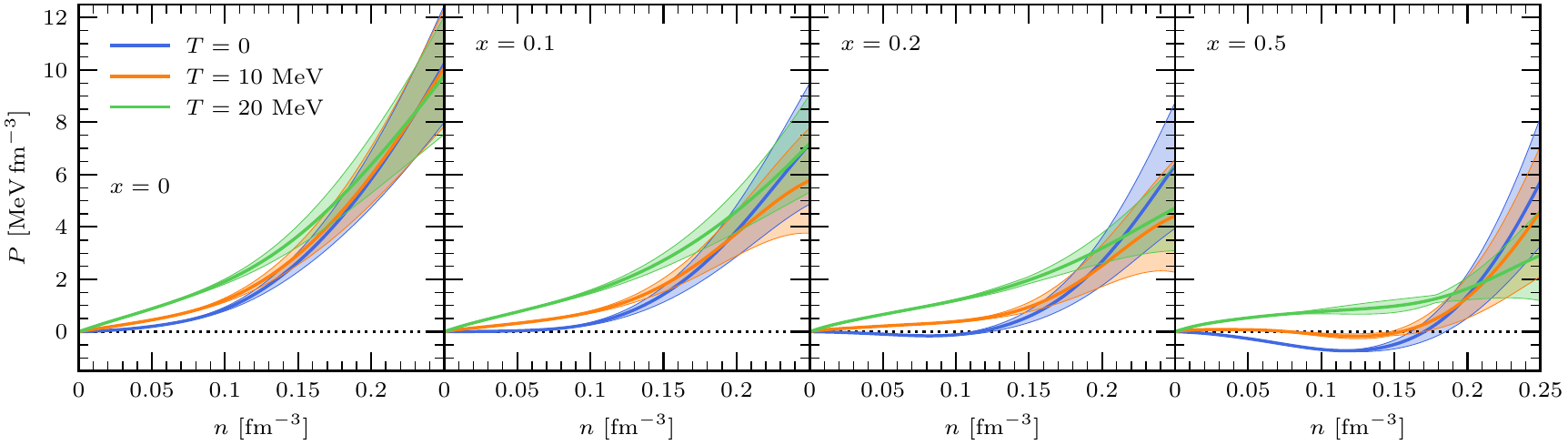}
    \caption{Same as Fig.~\ref{fig:free-energy} but for the pressure $P$ at N$^3$LO ($\Lambda = 450\mev$)
      from the GP emulator and with EFT uncertainties.}
    \label{fig:pressure}
\end{figure*}

For estimates of the theoretical uncertainties for an observable $X$ due to the truncated chiral expansion we use the prescription of
Ref.~\cite{EKM2015},
\begin{align}
  \Delta X^{(j)} &= Q \cdot\max\left(|X^{(j)} - X^{(j-1)}|, \Delta X^{(j-1)} \right) \, , \label{eq:ekm}
\end{align}
where $X^{(j)}$ is the observable calculated at
N$^j$LO and the expansion parameter is $Q = p / \Lambda_b$, where
we take $\Lambda_b = 500\,\text{MeV}$ for the EFT breakdown scale and
$p$ is a typical momentum for the observable of interest. We take $p$ to be the root-mean-square
momentum of the Fermi gas
$p^2 = \langle k^2 \rangle = 3 T (\sum_\tau m_\tau^{5/2} F_{3/2}(\mu_\tau / T)) / (\sum_\tau m_\tau^{3/2} F_{1/2}(\mu_\tau / T))$, 
where the chemical potentials $\mu_\tau$ are determined from the density
$n_\tau = 2^{-1/2} \left(m_\tau T/ \pi\right)^{3/2} F_{1/2}(\mu_\tau / T)$
and $F_n(x) = \Gamma(n+1)^{-1}\int_0^\infty \mathrm{d}t\,t^n (1 + \exp(t - x))^{-1}$
are Fermi integrals. The resulting EFT uncertainty bands at N$^3$LO are shown for
the third-order MBPT results in Fig.~\ref{fig:free-energy}.
In addition, we show the first-order (HF) and second-order MBPT(2) results to
assess the MBPT convergence of the
expansion. Table~\ref{tab:mbpt-convergence} gives numerical
values at fiducial $n, x, T$ to document the MBPT and chiral convergence. Overall, 
we find a systematic MBPT convergence.
For the first-order liquid-gas phase transition in symmetric nuclear matter, our results give the preliminary ranges for the critical temperature, density, and pressure, $T_c = 15.9-16.3$\,MeV, $n_c = 0.07-0.11$\,fm$^{-3}$, and $P_c = 0.30-0.40$\,MeV\,fm$^{-3}$, where the ranges are obtained by considering the N$^3$LO interaction at MBPT(3) and MBPT(2). A full analysis will be the topic of future work.

\prlsection{Pressure and thermal effects} The pressure
$P = n^2 \partial_n (F / A)|_{x,T} = P_\text{FG} + n^2 \partial_n
(F_\text{int} / A)|_{x,T}$ is shown in Fig.~\ref{fig:pressure} for
different proton fractions and temperatures, where the derivative of
the interaction energy $F_\text{int}/A$ is calculated using the GP
emulator. As expected, the pressure decreases with increasing proton fraction,
and for very neutron-rich conditions depends only weakly on the temperature 
for $n \gtrsim n_0$. Interestingly, for symmetric matter we find that the pressure 
decreases with increasing temperature for $n \gtrsim 0.2\,\text{fm}^{-3}$.
This negative thermal expansion has also been observed in 
Ref.~\cite{PhysRevC.89.064009} for low-momentum interactions. For
neutron-rich matter, this behavior is seen in Fig.~\ref{fig:pressure}
starting at higher densities.

\begin{figure}[t]
    \centering
    \includegraphics[clip=,width=\columnwidth]{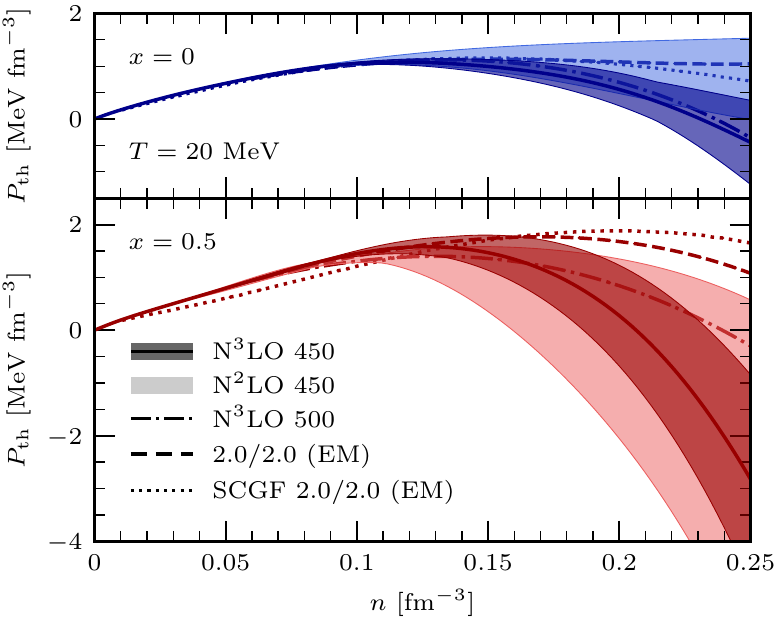}
    \caption{Thermal pressure $P_\text{th}$ for neutron matter (blue)
      and symmetric matter (red) for $T=20$\,MeV as a function of
      density. In addition to the N$^3$LO results with $\Lambda = 450\mev$ 
      (solid lines) we also show $\Lambda = 500\mev$ (dot-dashed lines) as well as
      for the 2.0/2.0 (EM) interaction~\cite{Hebeler_et_al2011} (dashed lines).
      For the latter, we compare against the SCGF results from 
      Ref.~\cite{CarboneSchwenk2019} (dotted lines).
      The darker EFT uncertainty bands are N$^3$LO, while the lighter ones are
      for N$^2$LO.}
    \label{fig:thermal-pressure}
\end{figure}

To investigate this further, we show the thermal pressure
$P_\mathrm{th} = P(T) - P(T=0)$ in Fig.~\ref{fig:thermal-pressure} for
neutron matter and symmetric matter for
$T = 20\,\text{MeV}$. We find that the thermal pressure starts to
decrease at $n \approx 0.15\,\text{fm}^{-3}$ and
becomes negative around $n \approx 0.2\,\text{fm}^{-3}$.
For neutron matter this finding is consistent with Ref.~\cite{Keller2021} 
and can be understood in terms of a neutron effective mass $m_n^*$ that
increases at higher density due to repulsive 3N contributions~\cite{CarboneSchwenk2019, Keller2021} 
[$P_\text{th} \leqslant 0$ requires $\partial m_n^*/\partial n_n \geqslant 0$,
see Eqs.~(39)~and~(41) in Ref.~\cite{Keller2021}].
Figure~\ref{fig:thermal-pressure} shows that a decreasing thermal pressure
at higher densities is found at different orders (N$^2$LO and N$^3$LO), different
cutoffs ($\Lambda = 450\mev$ and $500\mev$), as well as for the 2.0/2.0 (EM) interaction~\cite{Hebeler_et_al2011}, while the size of the decrease has large
theoretical uncertainties. For the 2.0/2.0 (EM) interaction, we can also
compare our MBPT against self-consistent Green's function (SCGF) results~\cite{CarboneSchwenk2019} and find good agreement (with the small differences likely due to the $T=0$ extrapolation and the normal-ordering approximation in Ref.~\cite{CarboneSchwenk2019}).
Note that the cutoff dependence of the negative thermal expansion might indicate that the maximal density accessible is limited based on the employed interactions.

\begin{figure}[t]
    \centering
    \includegraphics[clip=,width=\columnwidth]{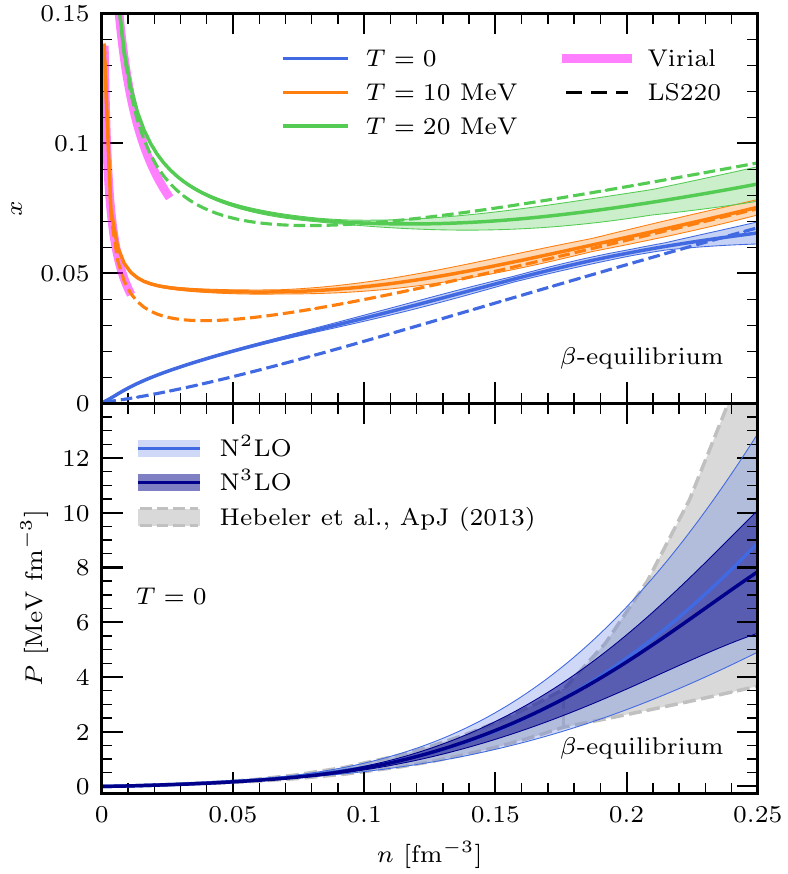}
    \caption{Upper panel: Proton fraction $x$ in beta equilibrium at N$^3$LO
      for different $T$ as a function of density. For comparison, we show the virial EOS
      and the LS220 EOS~\cite{Lattimer_Swesty_1991}. Lower panel: Pressure $P$ in beta 
      equilibrium for $T=0$ at N$^2$LO and N$^3$LO. We compare against the EOS band
      from Hebeler \textit{et al.}~\cite{Hebeler2013} based on chiral EFT interactions up
      to $1.1 n_0$ 
      and a general piecewise polytrope extension to higher densities.
      }
    \label{fig:beta-eq-x-P}
\end{figure}

\prlsection{Matter in beta equilibrium} 
Using the GP, we can access arbitrary proton fractions and derive other quantities through thermodynamically consistent derivatives. We first use the GP to calculate the proton fraction $x$ of neutron star matter in beta equilibrium as a function of density for different temperatures. For a given density and temperature, $x$ is determined
by the condition $m_n + \mu_n = m_p + \mu_p + m_e + \mu_e$, where the
neutron and proton chemical potentials are given by
$\mu_\tau = F/A + n \partial_n (F/A) + \left(\delta_{\tau,p} - x
\right) \partial_x (F/A)$. The electron chemical potential is
determined from the density of an ultra-relativistic Fermi gas,
$n_e = 2/\pi^2 T^3 F_2(\mu_e/T)$ with the Fermi integral
$F_2$ through charge neutrality $n_p = n_e$. Our results using the GP emulator are shown
in the upper panel of Fig.~\ref{fig:beta-eq-x-P}. We find very narrow
EFT uncertainty bands in this case, using again Eq.~(\ref{eq:ekm}) with
$Q = Q(n, x=x^\text{N$^3$LO}_{\beta\text{-eq.}}(n,T), T)$. At small densities 
and finite $T$, the proton fraction is dominated by the kinetic
part and follows the virial EOS. At higher densities,
the density dependence of $x$ is weaker, with proton fractions of
$4-8$\,\% for the temperatures considered. Overall, we find a reasonable
agreement with the Lattimer and Swesty EOS LS220~\cite{Lattimer_Swesty_1991}
but our chiral EFT results exhibit a weaker density dependence.

\begin{figure}[t]
    \centering
    \includegraphics[clip=,width=\columnwidth]{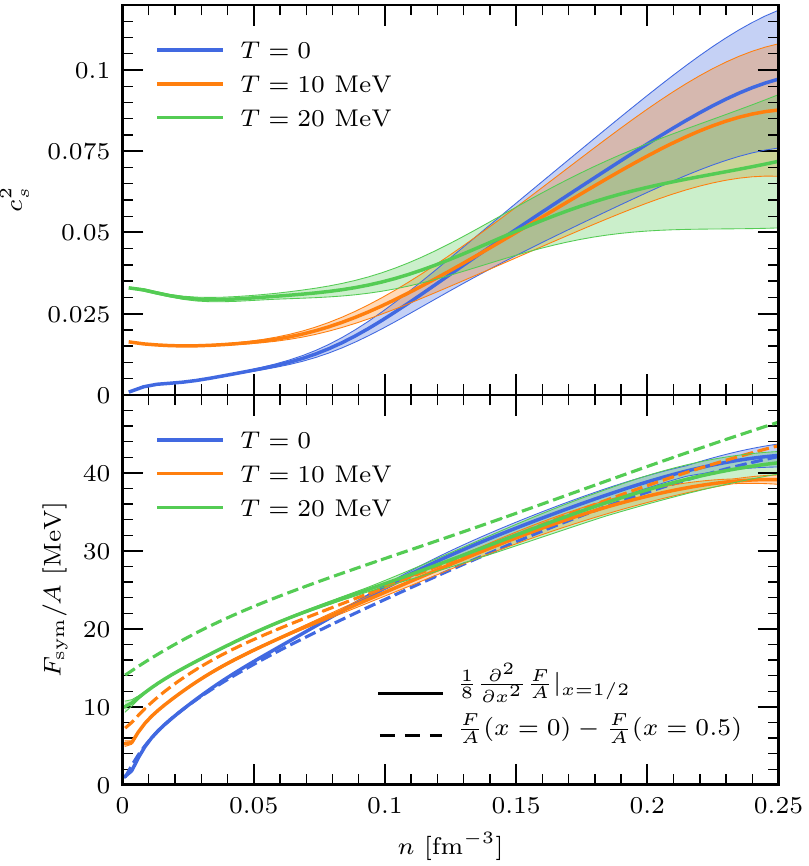}
    \caption{Speed of sound squared $c_s^2$ for neutron matter (upper panel) and 
      symmetry free energy per particle $F_\text{sym}/A$ (lower panel) at N$^3$LO for different $T$ as a function
      of density. We show results based on different definitions for
      $F_\text{sym}/A$ (second derivative around symmetric matter and difference
      between neutron and symmetric matter).
      }
    \label{fig:symmetry-energy}
\end{figure}

The GP thus also enables a first nonparametric calculation of the neutron star
EOS. In the lower panel of Fig.~\ref{fig:beta-eq-x-P}
we show the pressure of matter in beta equilibrium $P(n, x_{\beta\text{-eq.}}, T=0)$ at N$^2$LO and N$^3$LO 
with EFT uncertainty estimates. The N$^3$LO band is systematically
smaller and overlaps with the N$^2$LO band over the full density range.
Moreover, both bands behave naturally towards higher densities and hence 
show no indication for a breakdown of the chiral expansion up to $n \lesssim
0.25\,\text{fm}^{-3}$.
For comparison, we also show the EOS band from 
Hebeler \textit{et al.}~\cite{Hebeler2013} based on chiral EFT interactions up
to $1.1 n_0$ and a general piecewise polytrope extension to higher densities 
constrained by causality and the observation of two-solar-mass neutron stars. 
This EOS band results mainly from variations of the chiral 3N forces, so that the comparison is not direct. 
Nevertheless, the overlap with the nonparametric N$^2$LO and N$^3$LO bands
is remarkable. Up to $n_0$, the N$^3$LO band prefers high
pressures in the Hebeler \textit{et al.} band~\cite{Hebeler2013} and at higher
$n$, it provides important new constraints.

\prlsection{Speed of sound and symmetry energy}
Next, we study the speed of sound $c_s^2 = \partial P / \partial\varepsilon|_{S,x} = \frac{n}{P + \varepsilon} \partial P / \partial n|_{S,x}$ with the internal energy density $\varepsilon = n (E/A + m_n)$. The derivative at constant entropy $S$ and constant proton fraction $x$ is performed numerically based on $P(n, x, T(n, x, S))$. Our results using the GP emulator are shown for neutron matter in the upper panel of Fig.~\ref{fig:symmetry-energy}. Given that $c_s^2$ is a second derivative, the EFT uncertainties are larger in this case. At $T=0$, $c_s^2$ increases monotonously while the increase is weaker at finite $T$. As for the pressure, $c_s^2$ decreases at higher densities with increasing $T$.

As another GP application, we show the symmetry free energy per 
particle $F_\text{sym}/A$ as a function 
of density for different $T$ in the the lower panel of 
Fig.~(\ref{fig:symmetry-energy}). We compare two common definitions: 
the second derivative around symmetric matter and the difference
between neutron matter and symmetric matter, where the difference probes the size 
of contributions beyond a quadratic $x$ dependence.
Since the numerical uncertainties are enhanced in second derivatives due to residual noise from the Monte Carlo integration, we calculate $\partial^2 (F/A) / \partial x^2 |_{x=1/2}$ by fitting a GP to each MBPT diagram individually.
The difference definition
is more sensitive to thermal effects due to the nonquadratic contributions
contained in the kinetic part. For the second-derivative definition, 
we find that $F_\text{sym}/A$ is narrowly predicted at N$^3$LO
at a fixed saturation density $n_0$, with $F_\text{sym}(n_0)/A \approx 30\mev$ 
at $T=0$, while the uncertainty increases if one allows 
this reference density to vary~\cite{Drischler2020}. Moreover, the
$T$ dependence of the symmetry energy is mild at $n_0$, with
larger increases at lower and higher $n$. 

\prlsection{Conclusions and outlook} 
We presented first microscopic calculations of the EOS at arbitrary proton 
fractions and finite temperature based on chiral NN and 3N interactions to 
N$^3$LO, including uncertainty estimates from the many-body 
calculation and the chiral expansion. For this an emulator of the interaction free energy per particle was constructed using Gaussian processes. We demonstrated
that this enables an efficient and accurate evaluation of the EOS and thermodynamic 
derivatives for arbitrary values of $n$, $x$, and $T$, where we considered the 
ranges $n \leqslant 0.25\,\text{fm}^{-3}$, $x \leqslant 0.5$, and $T \leqslant 
20\,\text{MeV}$. The EFT uncertainties dominate over the MBPT
uncertainties for these nuclear densities.

We studied in detail the dependence of the free energy and the pressure on
proton fraction and temperature, and found that the pressure at 
higher densities decreases with increasing temperature, thus exhibiting a 
negative thermal expansion. The GP emulator allowed us to calculate the EOS in beta 
equilibrium directly without parametrizations between neutron and symmetric matter.
The resulting N$^3$LO neutron star EOS exhibited a systematic chiral 
EFT behavior over the full range ($n \leqslant 0.25\,\text{fm}^{-3}$) and 
significantly improved the uncertainties over previous EOS bands, preferring larger
values for the pressure. Moreover, we presented first microscopic results 
for the speed of sound and the symmetry energy at finite temperature.
Our framework and results test commonly applied approximation for
the proton fraction and temperature dependence of the EOS and open the door 
to nonparametric EOS input for astrophysical simulations of
supernovae and mergers.

We thank P. Arthuis for providing benchmark expressions for $T=0$ MBPT diagrams. This work was supported by the Deutsche Forschungsgemeinschaft
(DFG, German Research Foundation) -- Project-ID 279384907 -- SFB 1245
and by the State of Hesse within the Research Cluster ELEMENTS 
(Project ID 500/10.006).

\bibliography{literature}

\begin{thebibliography}{51}%
\makeatletter
\providecommand \@ifxundefined [1]{%
 \@ifx{#1\undefined}
}%
\providecommand \@ifnum [1]{%
 \ifnum #1\expandafter \@firstoftwo
 \else \expandafter \@secondoftwo
 \fi
}%
\providecommand \@ifx [1]{%
 \ifx #1\expandafter \@firstoftwo
 \else \expandafter \@secondoftwo
 \fi
}%
\providecommand \natexlab [1]{#1}%
\providecommand \enquote  [1]{``#1''}%
\providecommand \bibnamefont  [1]{#1}%
\providecommand \bibfnamefont [1]{#1}%
\providecommand \citenamefont [1]{#1}%
\providecommand \href@noop [0]{\@secondoftwo}%
\providecommand \href [0]{\begingroup \@sanitize@url \@href}%
\providecommand \@href[1]{\@@startlink{#1}\@@href}%
\providecommand \@@href[1]{\endgroup#1\@@endlink}%
\providecommand \@sanitize@url [0]{\catcode `\\12\catcode `\$12\catcode
  `\&12\catcode `\#12\catcode `\^12\catcode `\_12\catcode `\%12\relax}%
\providecommand \@@startlink[1]{}%
\providecommand \@@endlink[0]{}%
\providecommand \url  [0]{\begingroup\@sanitize@url \@url }%
\providecommand \@url [1]{\endgroup\@href {#1}{\urlprefix }}%
\providecommand \urlprefix  [0]{URL }%
\providecommand \Eprint [0]{\href }%
\providecommand \doibase [0]{https://doi.org/}%
\providecommand \selectlanguage [0]{\@gobble}%
\providecommand \bibinfo  [0]{\@secondoftwo}%
\providecommand \bibfield  [0]{\@secondoftwo}%
\providecommand \translation [1]{[#1]}%
\providecommand \BibitemOpen [0]{}%
\providecommand \bibitemStop [0]{}%
\providecommand \bibitemNoStop [0]{.\EOS\space}%
\providecommand \EOS [0]{\spacefactor3000\relax}%
\providecommand \BibitemShut  [1]{\csname bibitem#1\endcsname}%
\let\auto@bib@innerbib\@empty
\bibitem [{\citenamefont {Lattimer}\ and\ \citenamefont
  {Lim}(2013)}]{Latt12esymm}%
  \BibitemOpen
  \bibfield  {author} {\bibinfo {author} {\bibfnamefont {J.~M.}\ \bibnamefont
  {Lattimer}}\ and\ \bibinfo {author} {\bibfnamefont {Y.}~\bibnamefont {Lim}},\
  }\bibfield  {title} {\bibinfo {title} {{Constraining the Symmetry Parameters
  of the Nuclear Interaction}},\ }\href
  {https://doi.org/10.1088/0004-637X/771/1/51} {\bibfield  {journal} {\bibinfo
  {journal} {Astrophys. J.}\ }\textbf {\bibinfo {volume} {771}},\ \bibinfo
  {pages} {51} (\bibinfo {year} {2013})}\BibitemShut {NoStop}%
\bibitem [{\citenamefont {Hebeler}\ \emph {et~al.}(2015)\citenamefont
  {Hebeler}, \citenamefont {Holt}, \citenamefont {Men\'endez},\ and\
  \citenamefont {Schwenk}}]{Hebeler2015}%
  \BibitemOpen
  \bibfield  {author} {\bibinfo {author} {\bibfnamefont {K.}~\bibnamefont
  {Hebeler}}, \bibinfo {author} {\bibfnamefont {J.~D.}\ \bibnamefont {Holt}},
  \bibinfo {author} {\bibfnamefont {J.}~\bibnamefont {Men\'endez}},\ and\
  \bibinfo {author} {\bibfnamefont {A.}~\bibnamefont {Schwenk}},\ }\bibfield
  {title} {\bibinfo {title} {{Nuclear forces and their impact on neutron-rich
  nuclei and neutron-rich matter}},\ }\href
  {https://doi.org/10.1146/annurev-nucl-102313-025446} {\bibfield  {journal}
  {\bibinfo  {journal} {Annu. Rev. Nucl. Part. Sci.}\ }\textbf {\bibinfo
  {volume} {65}},\ \bibinfo {pages} {457} (\bibinfo {year} {2015})}\BibitemShut
  {NoStop}%
\bibitem [{\citenamefont {Lynn}\ \emph {et~al.}(2019)\citenamefont {Lynn},
  \citenamefont {Tews}, \citenamefont {Gandolfi},\ and\ \citenamefont
  {Lovato}}]{Lynn2019}%
  \BibitemOpen
  \bibfield  {author} {\bibinfo {author} {\bibfnamefont {J.~E.}\ \bibnamefont
  {Lynn}}, \bibinfo {author} {\bibfnamefont {I.}~\bibnamefont {Tews}}, \bibinfo
  {author} {\bibfnamefont {S.}~\bibnamefont {Gandolfi}},\ and\ \bibinfo
  {author} {\bibfnamefont {A.}~\bibnamefont {Lovato}},\ }\bibfield  {title}
  {\bibinfo {title} {{Quantum Monte Carlo Methods in Nuclear Physics: Recent
  Advances}},\ }\href {https://doi.org/10.1146/annurev-nucl-101918-023600}
  {\bibfield  {journal} {\bibinfo  {journal} {Annu. Rev. Nucl. Part. Sci.}\
  }\textbf {\bibinfo {volume} {69}},\ \bibinfo {pages} {279} (\bibinfo {year}
  {2019})}\BibitemShut {NoStop}%
\bibitem [{\citenamefont {Drischler}\ \emph
  {et~al.}(2021{\natexlab{a}})\citenamefont {Drischler}, \citenamefont {Holt},\
  and\ \citenamefont {Wellenhofer}}]{Drischler2021ARNPS}%
  \BibitemOpen
  \bibfield  {author} {\bibinfo {author} {\bibfnamefont {C.}~\bibnamefont
  {Drischler}}, \bibinfo {author} {\bibfnamefont {J.~W.}\ \bibnamefont
  {Holt}},\ and\ \bibinfo {author} {\bibfnamefont {C.}~\bibnamefont
  {Wellenhofer}},\ }\bibfield  {title} {\bibinfo {title} {{Chiral Effective
  Field Theory and the High-Density Nuclear Equation of State}},\ }\href
  {https://doi.org/10.1146/annurev-nucl-102419-041903} {\bibfield  {journal}
  {\bibinfo  {journal} {Annu. Rev. Nucl. Part. Sci.}\ }\textbf {\bibinfo
  {volume} {71}},\ \bibinfo {pages} {403} (\bibinfo {year}
  {2021}{\natexlab{a}})}\BibitemShut {NoStop}%
\bibitem [{\citenamefont {Lattimer}(2021)}]{Lattimer2021}%
  \BibitemOpen
  \bibfield  {author} {\bibinfo {author} {\bibfnamefont {J.~M.}\ \bibnamefont
  {Lattimer}},\ }\bibfield  {title} {\bibinfo {title} {{Neutron Stars and the
  Nuclear Matter Equation of State}},\ }\href
  {https://doi.org/10.1146/annurev-nucl-102419-124827} {\bibfield  {journal}
  {\bibinfo  {journal} {Annu. Rev. Nucl. Part. Sci.}\ }\textbf {\bibinfo
  {volume} {71}},\ \bibinfo {pages} {433} (\bibinfo {year} {2021})}\BibitemShut
  {NoStop}%
\bibitem [{\citenamefont {Epelbaum}\ \emph {et~al.}(2009)\citenamefont
  {Epelbaum}, \citenamefont {Hammer},\ and\ \citenamefont
  {Mei\ss{}ner}}]{Epelbaum_et_al_2009}%
  \BibitemOpen
  \bibfield  {author} {\bibinfo {author} {\bibfnamefont {E.}~\bibnamefont
  {Epelbaum}}, \bibinfo {author} {\bibfnamefont {H.-W.}\ \bibnamefont
  {Hammer}},\ and\ \bibinfo {author} {\bibfnamefont {U.-G.}\ \bibnamefont
  {Mei\ss{}ner}},\ }\bibfield  {title} {\bibinfo {title} {Modern theory of
  nuclear forces},\ }\href {https://doi.org/10.1103/RevModPhys.81.1773}
  {\bibfield  {journal} {\bibinfo  {journal} {Rev. Mod. Phys.}\ }\textbf
  {\bibinfo {volume} {81}},\ \bibinfo {pages} {1773} (\bibinfo {year}
  {2009})}\BibitemShut {NoStop}%
\bibitem [{\citenamefont {Machleidt}\ and\ \citenamefont
  {Entem}(2011)}]{Machleidt_Entem_2011}%
  \BibitemOpen
  \bibfield  {author} {\bibinfo {author} {\bibfnamefont {R.}~\bibnamefont
  {Machleidt}}\ and\ \bibinfo {author} {\bibfnamefont {D.~R.}\ \bibnamefont
  {Entem}},\ }\bibfield  {title} {\bibinfo {title} {Chiral effective field
  theory and nuclear forces},\ }\href
  {https://doi.org/https://doi.org/10.1016/j.physrep.2011.02.001} {\bibfield
  {journal} {\bibinfo  {journal} {Phys. Rept.}\ }\textbf {\bibinfo {volume}
  {503}},\ \bibinfo {pages} {1} (\bibinfo {year} {2011})}\BibitemShut {NoStop}%
\bibitem [{\citenamefont {Hammer}\ \emph {et~al.}(2020)\citenamefont {Hammer},
  \citenamefont {K\"onig},\ and\ \citenamefont {van Kolck}}]{Hammer2020}%
  \BibitemOpen
  \bibfield  {author} {\bibinfo {author} {\bibfnamefont {H.-W.}\ \bibnamefont
  {Hammer}}, \bibinfo {author} {\bibfnamefont {S.}~\bibnamefont {K\"onig}},\
  and\ \bibinfo {author} {\bibfnamefont {U.}~\bibnamefont {van Kolck}},\
  }\bibfield  {title} {\bibinfo {title} {Nuclear effective field theory: Status
  and perspectives},\ }\href {https://doi.org/10.1103/RevModPhys.92.025004}
  {\bibfield  {journal} {\bibinfo  {journal} {Rev. Mod. Phys.}\ }\textbf
  {\bibinfo {volume} {92}},\ \bibinfo {pages} {025004} (\bibinfo {year}
  {2020})}\BibitemShut {NoStop}%
\bibitem [{\citenamefont {Hebeler}\ and\ \citenamefont
  {Schwenk}(2010)}]{HebelerSchwenk2010}%
  \BibitemOpen
  \bibfield  {author} {\bibinfo {author} {\bibfnamefont {K.}~\bibnamefont
  {Hebeler}}\ and\ \bibinfo {author} {\bibfnamefont {A.}~\bibnamefont
  {Schwenk}},\ }\bibfield  {title} {\bibinfo {title} {Chiral three-nucleon
  forces and neutron matter},\ }\href
  {https://doi.org/10.1103/PhysRevC.82.014314} {\bibfield  {journal} {\bibinfo
  {journal} {Phys. Rev. C}\ }\textbf {\bibinfo {volume} {82}},\ \bibinfo
  {pages} {014314} (\bibinfo {year} {2010})}\BibitemShut {NoStop}%
\bibitem [{\citenamefont {Hebeler}\ \emph {et~al.}(2011)\citenamefont
  {Hebeler}, \citenamefont {Bogner}, \citenamefont {Furnstahl}, \citenamefont
  {Nogga},\ and\ \citenamefont {Schwenk}}]{Hebeler_et_al2011}%
  \BibitemOpen
  \bibfield  {author} {\bibinfo {author} {\bibfnamefont {K.}~\bibnamefont
  {Hebeler}}, \bibinfo {author} {\bibfnamefont {S.~K.}\ \bibnamefont {Bogner}},
  \bibinfo {author} {\bibfnamefont {R.~J.}\ \bibnamefont {Furnstahl}}, \bibinfo
  {author} {\bibfnamefont {A.}~\bibnamefont {Nogga}},\ and\ \bibinfo {author}
  {\bibfnamefont {A.}~\bibnamefont {Schwenk}},\ }\bibfield  {title} {\bibinfo
  {title} {Improved nuclear matter calculations from chiral low-momentum
  interactions},\ }\href {https://doi.org/10.1103/PhysRevC.83.031301}
  {\bibfield  {journal} {\bibinfo  {journal} {Phys. Rev. C}\ }\textbf {\bibinfo
  {volume} {83}},\ \bibinfo {pages} {031301(R)} (\bibinfo {year}
  {2011})}\BibitemShut {NoStop}%
\bibitem [{\citenamefont {Tews}\ \emph {et~al.}(2013)\citenamefont {Tews},
  \citenamefont {Kr{\"u}ger}, \citenamefont {Hebeler},\ and\ \citenamefont
  {Schwenk}}]{Tews13N3LO}%
  \BibitemOpen
  \bibfield  {author} {\bibinfo {author} {\bibfnamefont {I.}~\bibnamefont
  {Tews}}, \bibinfo {author} {\bibfnamefont {T.}~\bibnamefont {Kr{\"u}ger}},
  \bibinfo {author} {\bibfnamefont {K.}~\bibnamefont {Hebeler}},\ and\ \bibinfo
  {author} {\bibfnamefont {A.}~\bibnamefont {Schwenk}},\ }\bibfield  {title}
  {\bibinfo {title} {{Neutron Matter at Next-to-Next-to-Next-to-Leading Order
  in Chiral Effective Field Theory}},\ }\href
  {https://doi.org/10.1103/PhysRevLett.110.032504} {\bibfield  {journal}
  {\bibinfo  {journal} {Phys. Rev. Lett.}\ }\textbf {\bibinfo {volume} {110}},\
  \bibinfo {pages} {032504} (\bibinfo {year} {2013})}\BibitemShut {NoStop}%
\bibitem [{\citenamefont {Holt}\ \emph {et~al.}(2013)\citenamefont {Holt},
  \citenamefont {Kaiser},\ and\ \citenamefont {Weise}}]{Holt13PPNP}%
  \BibitemOpen
  \bibfield  {author} {\bibinfo {author} {\bibfnamefont {J.~W.}\ \bibnamefont
  {Holt}}, \bibinfo {author} {\bibfnamefont {N.}~\bibnamefont {Kaiser}},\ and\
  \bibinfo {author} {\bibfnamefont {W.}~\bibnamefont {Weise}},\ }\bibfield
  {title} {\bibinfo {title} {{Nuclear chiral dynamics and thermodynamics}},\
  }\href {https://doi.org/10.1016/j.ppnp.2013.08.001} {\bibfield  {journal}
  {\bibinfo  {journal} {Prog. Part. Nucl. Phys.}\ }\textbf {\bibinfo {volume}
  {73}},\ \bibinfo {pages} {35} (\bibinfo {year} {2013})}\BibitemShut {NoStop}%
\bibitem [{\citenamefont {Carbone}\ \emph {et~al.}(2013)\citenamefont
  {Carbone}, \citenamefont {Polls},\ and\ \citenamefont {Rios}}]{Carb13nm}%
  \BibitemOpen
  \bibfield  {author} {\bibinfo {author} {\bibfnamefont {A.}~\bibnamefont
  {Carbone}}, \bibinfo {author} {\bibfnamefont {A.}~\bibnamefont {Polls}},\
  and\ \bibinfo {author} {\bibfnamefont {A.}~\bibnamefont {Rios}},\ }\bibfield
  {title} {\bibinfo {title} {{Symmetric nuclear matter with chiral
  three-nucleon forces in the self-consistent Green's functions approach}},\
  }\href {https://doi.org/10.1103/PhysRevC.88.044302} {\bibfield  {journal}
  {\bibinfo  {journal} {Phys. Rev. C}\ }\textbf {\bibinfo {volume} {88}},\
  \bibinfo {pages} {044302} (\bibinfo {year} {2013})}\BibitemShut {NoStop}%
\bibitem [{\citenamefont {Hagen}\ \emph {et~al.}(2014)\citenamefont {Hagen},
  \citenamefont {Papenbrock}, \citenamefont {Ekstr{\"o}m}, \citenamefont
  {Wendt}, \citenamefont {Baardsen}, \citenamefont {Gandolfi}, \citenamefont
  {Hjorth-Jensen},\ and\ \citenamefont {Horowitz}}]{Hage14ccnm}%
  \BibitemOpen
  \bibfield  {author} {\bibinfo {author} {\bibfnamefont {G.}~\bibnamefont
  {Hagen}}, \bibinfo {author} {\bibfnamefont {T.}~\bibnamefont {Papenbrock}},
  \bibinfo {author} {\bibfnamefont {A.}~\bibnamefont {Ekstr{\"o}m}}, \bibinfo
  {author} {\bibfnamefont {K.}~\bibnamefont {Wendt}}, \bibinfo {author}
  {\bibfnamefont {G.}~\bibnamefont {Baardsen}}, \bibinfo {author}
  {\bibfnamefont {S.}~\bibnamefont {Gandolfi}}, \bibinfo {author}
  {\bibfnamefont {M.}~\bibnamefont {Hjorth-Jensen}},\ and\ \bibinfo {author}
  {\bibfnamefont {C.~J.}\ \bibnamefont {Horowitz}},\ }\bibfield  {title}
  {\bibinfo {title} {{Coupled-cluster calculations of nucleonic matter}},\
  }\href {https://doi.org/10.1103/PhysRevC.89.014319} {\bibfield  {journal}
  {\bibinfo  {journal} {Phys. Rev. C}\ }\textbf {\bibinfo {volume} {89}},\
  \bibinfo {pages} {014319} (\bibinfo {year} {2014})}\BibitemShut {NoStop}%
\bibitem [{\citenamefont {Coraggio}\ \emph {et~al.}(2014)\citenamefont
  {Coraggio}, \citenamefont {Holt}, \citenamefont {Itaco}, \citenamefont
  {Machleidt}, \citenamefont {Marcucci},\ and\ \citenamefont
  {Sammarruca}}]{Coraggio_et_al_2014}%
  \BibitemOpen
  \bibfield  {author} {\bibinfo {author} {\bibfnamefont {L.}~\bibnamefont
  {Coraggio}}, \bibinfo {author} {\bibfnamefont {J.~W.}\ \bibnamefont {Holt}},
  \bibinfo {author} {\bibfnamefont {N.}~\bibnamefont {Itaco}}, \bibinfo
  {author} {\bibfnamefont {R.}~\bibnamefont {Machleidt}}, \bibinfo {author}
  {\bibfnamefont {L.~E.}\ \bibnamefont {Marcucci}},\ and\ \bibinfo {author}
  {\bibfnamefont {F.}~\bibnamefont {Sammarruca}},\ }\bibfield  {title}
  {\bibinfo {title} {Nuclear-matter equation of state with consistent two- and
  three-body perturbative chiral interactions},\ }\href
  {https://doi.org/10.1103/PhysRevC.89.044321} {\bibfield  {journal} {\bibinfo
  {journal} {Phys. Rev. C}\ }\textbf {\bibinfo {volume} {89}},\ \bibinfo
  {pages} {044321} (\bibinfo {year} {2014})}\BibitemShut {NoStop}%
\bibitem [{\citenamefont {Wellenhofer}\ \emph {et~al.}(2014)\citenamefont
  {Wellenhofer}, \citenamefont {Holt}, \citenamefont {Kaiser},\ and\
  \citenamefont {Weise}}]{PhysRevC.89.064009}%
  \BibitemOpen
  \bibfield  {author} {\bibinfo {author} {\bibfnamefont {C.}~\bibnamefont
  {Wellenhofer}}, \bibinfo {author} {\bibfnamefont {J.~W.}\ \bibnamefont
  {Holt}}, \bibinfo {author} {\bibfnamefont {N.}~\bibnamefont {Kaiser}},\ and\
  \bibinfo {author} {\bibfnamefont {W.}~\bibnamefont {Weise}},\ }\bibfield
  {title} {\bibinfo {title} {Nuclear thermodynamics from chiral low-momentum
  interactions},\ }\href {https://doi.org/10.1103/PhysRevC.89.064009}
  {\bibfield  {journal} {\bibinfo  {journal} {Phys. Rev. C}\ }\textbf {\bibinfo
  {volume} {89}},\ \bibinfo {pages} {064009} (\bibinfo {year}
  {2014})}\BibitemShut {NoStop}%
\bibitem [{\citenamefont {Wellenhofer}\ \emph {et~al.}(2015)\citenamefont
  {Wellenhofer}, \citenamefont {Holt},\ and\ \citenamefont
  {Kaiser}}]{PhysRevC.92.015801}%
  \BibitemOpen
  \bibfield  {author} {\bibinfo {author} {\bibfnamefont {C.}~\bibnamefont
  {Wellenhofer}}, \bibinfo {author} {\bibfnamefont {J.~W.}\ \bibnamefont
  {Holt}},\ and\ \bibinfo {author} {\bibfnamefont {N.}~\bibnamefont {Kaiser}},\
  }\bibfield  {title} {\bibinfo {title} {Thermodynamics of isospin-asymmetric
  nuclear matter from chiral effective field theory},\ }\href
  {https://doi.org/10.1103/PhysRevC.92.015801} {\bibfield  {journal} {\bibinfo
  {journal} {Phys. Rev. C}\ }\textbf {\bibinfo {volume} {92}},\ \bibinfo
  {pages} {015801} (\bibinfo {year} {2015})}\BibitemShut {NoStop}%
\bibitem [{\citenamefont {Lynn}\ \emph {et~al.}(2016)\citenamefont {Lynn},
  \citenamefont {Tews}, \citenamefont {Carlson}, \citenamefont {Gandolfi},
  \citenamefont {Gezerlis}, \citenamefont {Schmidt},\ and\ \citenamefont
  {Schwenk}}]{Lynn16QMC3N}%
  \BibitemOpen
  \bibfield  {author} {\bibinfo {author} {\bibfnamefont {J.~E.}\ \bibnamefont
  {Lynn}}, \bibinfo {author} {\bibfnamefont {I.}~\bibnamefont {Tews}}, \bibinfo
  {author} {\bibfnamefont {J.}~\bibnamefont {Carlson}}, \bibinfo {author}
  {\bibfnamefont {S.}~\bibnamefont {Gandolfi}}, \bibinfo {author}
  {\bibfnamefont {A.}~\bibnamefont {Gezerlis}}, \bibinfo {author}
  {\bibfnamefont {K.~E.}\ \bibnamefont {Schmidt}},\ and\ \bibinfo {author}
  {\bibfnamefont {A.}~\bibnamefont {Schwenk}},\ }\bibfield  {title} {\bibinfo
  {title} {{Chiral Three-Nucleon Interactions in Light Nuclei, Neutron-$\alpha$
  Scattering, and Neutron Matter}},\ }\href
  {https://doi.org/10.1103/PhysRevLett.116.062501} {\bibfield  {journal}
  {\bibinfo  {journal} {Phys. Rev. Lett.}\ }\textbf {\bibinfo {volume} {116}},\
  \bibinfo {pages} {062501} (\bibinfo {year} {2016})}\BibitemShut {NoStop}%
\bibitem [{\citenamefont {Drischler}\ \emph {et~al.}(2016)\citenamefont
  {Drischler}, \citenamefont {Hebeler},\ and\ \citenamefont
  {Schwenk}}]{Dris16asym}%
  \BibitemOpen
  \bibfield  {author} {\bibinfo {author} {\bibfnamefont {C.}~\bibnamefont
  {Drischler}}, \bibinfo {author} {\bibfnamefont {K.}~\bibnamefont {Hebeler}},\
  and\ \bibinfo {author} {\bibfnamefont {A.}~\bibnamefont {Schwenk}},\
  }\bibfield  {title} {\bibinfo {title} {{Asymmetric nuclear matter based on
  chiral two- and three-nucleon interactions}},\ }\href
  {https://doi.org/10.1103/PhysRevC.93.054314} {\bibfield  {journal} {\bibinfo
  {journal} {Phys. Rev. C}\ }\textbf {\bibinfo {volume} {93}},\ \bibinfo
  {pages} {054314} (\bibinfo {year} {2016})}\BibitemShut {NoStop}%
\bibitem [{\citenamefont {Ekstr{\"o}m}\ \emph {et~al.}(2018)\citenamefont
  {Ekstr{\"o}m}, \citenamefont {Hagen}, \citenamefont {Morris}, \citenamefont
  {Papenbrock},\ and\ \citenamefont {Schwartz}}]{Ekst17deltasat}%
  \BibitemOpen
  \bibfield  {author} {\bibinfo {author} {\bibfnamefont {A.}~\bibnamefont
  {Ekstr{\"o}m}}, \bibinfo {author} {\bibfnamefont {G.}~\bibnamefont {Hagen}},
  \bibinfo {author} {\bibfnamefont {T.~D.}\ \bibnamefont {Morris}}, \bibinfo
  {author} {\bibfnamefont {T.}~\bibnamefont {Papenbrock}},\ and\ \bibinfo
  {author} {\bibfnamefont {P.~D.}\ \bibnamefont {Schwartz}},\ }\bibfield
  {title} {\bibinfo {title} {{$\Delta$ isobars and nuclear saturation}},\
  }\href {https://doi.org/10.1103/PhysRevC.97.024332} {\bibfield  {journal}
  {\bibinfo  {journal} {Phys. Rev. C}\ }\textbf {\bibinfo {volume} {97}},\
  \bibinfo {pages} {024332} (\bibinfo {year} {2018})}\BibitemShut {NoStop}%
\bibitem [{\citenamefont {Carbone}\ \emph {et~al.}(2018)\citenamefont
  {Carbone}, \citenamefont {Polls},\ and\ \citenamefont
  {Rios}}]{Carbone_et_al_2018}%
  \BibitemOpen
  \bibfield  {author} {\bibinfo {author} {\bibfnamefont {A.}~\bibnamefont
  {Carbone}}, \bibinfo {author} {\bibfnamefont {A.}~\bibnamefont {Polls}},\
  and\ \bibinfo {author} {\bibfnamefont {A.}~\bibnamefont {Rios}},\ }\bibfield
  {title} {\bibinfo {title} {{Microscopic Predictions of the Nuclear Matter
  Liquid-Gas Phase Transition}},\ }\href
  {https://doi.org/10.1103/PhysRevC.98.025804} {\bibfield  {journal} {\bibinfo
  {journal} {Phys. Rev. C}\ }\textbf {\bibinfo {volume} {98}},\ \bibinfo
  {pages} {025804} (\bibinfo {year} {2018})}\BibitemShut {NoStop}%
\bibitem [{\citenamefont {Drischler}\ \emph {et~al.}(2019)\citenamefont
  {Drischler}, \citenamefont {Hebeler},\ and\ \citenamefont
  {Schwenk}}]{Drischler2019}%
  \BibitemOpen
  \bibfield  {author} {\bibinfo {author} {\bibfnamefont {C.}~\bibnamefont
  {Drischler}}, \bibinfo {author} {\bibfnamefont {K.}~\bibnamefont {Hebeler}},\
  and\ \bibinfo {author} {\bibfnamefont {A.}~\bibnamefont {Schwenk}},\
  }\bibfield  {title} {\bibinfo {title} {Chiral interactions up to
  next-to-next-to-next-to-leading order and nuclear saturation},\ }\href
  {https://doi.org/10.1103/PhysRevLett.122.042501} {\bibfield  {journal}
  {\bibinfo  {journal} {Phys. Rev. Lett.}\ }\textbf {\bibinfo {volume} {122}},\
  \bibinfo {pages} {042501} (\bibinfo {year} {2019})}\BibitemShut {NoStop}%
\bibitem [{\citenamefont {Carbone}\ and\ \citenamefont
  {Schwenk}(2019)}]{CarboneSchwenk2019}%
  \BibitemOpen
  \bibfield  {author} {\bibinfo {author} {\bibfnamefont {A.}~\bibnamefont
  {Carbone}}\ and\ \bibinfo {author} {\bibfnamefont {A.}~\bibnamefont
  {Schwenk}},\ }\bibfield  {title} {\bibinfo {title} {Ab initio constraints on
  thermal effects of the nuclear equation of state},\ }\href
  {https://doi.org/10.1103/PhysRevC.100.025805} {\bibfield  {journal} {\bibinfo
   {journal} {Phys. Rev. C}\ }\textbf {\bibinfo {volume} {100}},\ \bibinfo
  {pages} {025805} (\bibinfo {year} {2019})}\BibitemShut {NoStop}%
\bibitem [{\citenamefont {Lu}\ \emph {et~al.}(2020)\citenamefont {Lu},
  \citenamefont {Li}, \citenamefont {Elhatisari}, \citenamefont {Lee},
  \citenamefont {Drut}, \citenamefont {L\"ahde}, \citenamefont {Epelbaum},\
  and\ \citenamefont {Mei\ss{}ner}}]{Lu2020}%
  \BibitemOpen
  \bibfield  {author} {\bibinfo {author} {\bibfnamefont {B.-N.}\ \bibnamefont
  {Lu}}, \bibinfo {author} {\bibfnamefont {N.}~\bibnamefont {Li}}, \bibinfo
  {author} {\bibfnamefont {S.}~\bibnamefont {Elhatisari}}, \bibinfo {author}
  {\bibfnamefont {D.}~\bibnamefont {Lee}}, \bibinfo {author} {\bibfnamefont
  {J.~E.}\ \bibnamefont {Drut}}, \bibinfo {author} {\bibfnamefont {T.~A.}\
  \bibnamefont {L\"ahde}}, \bibinfo {author} {\bibfnamefont {E.}~\bibnamefont
  {Epelbaum}},\ and\ \bibinfo {author} {\bibfnamefont {U.-G.}\ \bibnamefont
  {Mei\ss{}ner}},\ }\bibfield  {title} {\bibinfo {title} {{Ab Initio Nuclear
  Thermodynamics}},\ }\href {https://doi.org/10.1103/PhysRevLett.125.192502}
  {\bibfield  {journal} {\bibinfo  {journal} {Phys. Rev. Lett.}\ }\textbf
  {\bibinfo {volume} {125}},\ \bibinfo {pages} {192502} (\bibinfo {year}
  {2020})}\BibitemShut {NoStop}%
\bibitem [{\citenamefont {Drischler}\ \emph
  {et~al.}(2020{\natexlab{a}})\citenamefont {Drischler}, \citenamefont
  {Furnstahl}, \citenamefont {Melendez},\ and\ \citenamefont
  {Phillips}}]{Drischler2020PRL}%
  \BibitemOpen
  \bibfield  {author} {\bibinfo {author} {\bibfnamefont {C.}~\bibnamefont
  {Drischler}}, \bibinfo {author} {\bibfnamefont {R.~J.}\ \bibnamefont
  {Furnstahl}}, \bibinfo {author} {\bibfnamefont {J.~A.}\ \bibnamefont
  {Melendez}},\ and\ \bibinfo {author} {\bibfnamefont {D.~R.}\ \bibnamefont
  {Phillips}},\ }\bibfield  {title} {\bibinfo {title} {{How Well Do We Know the
  Neutron-Matter Equation of State at the Densities Inside Neutron Stars? A
  Bayesian Approach with Correlated Uncertainties}},\ }\href
  {https://doi.org/10.1103/PhysRevLett.125.202702} {\bibfield  {journal}
  {\bibinfo  {journal} {Phys. Rev. Lett.}\ }\textbf {\bibinfo {volume} {125}},\
  \bibinfo {pages} {202702} (\bibinfo {year} {2020}{\natexlab{a}})}\BibitemShut
  {NoStop}%
\bibitem [{\citenamefont {Keller}\ \emph {et~al.}(2021)\citenamefont {Keller},
  \citenamefont {Wellenhofer}, \citenamefont {Hebeler},\ and\ \citenamefont
  {Schwenk}}]{Keller2021}%
  \BibitemOpen
  \bibfield  {author} {\bibinfo {author} {\bibfnamefont {J.}~\bibnamefont
  {Keller}}, \bibinfo {author} {\bibfnamefont {C.}~\bibnamefont {Wellenhofer}},
  \bibinfo {author} {\bibfnamefont {K.}~\bibnamefont {Hebeler}},\ and\ \bibinfo
  {author} {\bibfnamefont {A.}~\bibnamefont {Schwenk}},\ }\bibfield  {title}
  {\bibinfo {title} {{Neutron matter at finite temperature based on chiral
  effective field theory interactions}},\ }\href
  {https://doi.org/10.1103/PhysRevC.103.055806} {\bibfield  {journal} {\bibinfo
   {journal} {Phys. Rev. C}\ }\textbf {\bibinfo {volume} {103}},\ \bibinfo
  {pages} {055806} (\bibinfo {year} {2021})}\BibitemShut {NoStop}%
\bibitem [{\citenamefont {Hebeler}\ \emph {et~al.}(2013)\citenamefont
  {Hebeler}, \citenamefont {Lattimer}, \citenamefont {Pethick},\ and\
  \citenamefont {Schwenk}}]{Hebeler2013}%
  \BibitemOpen
  \bibfield  {author} {\bibinfo {author} {\bibfnamefont {K.}~\bibnamefont
  {Hebeler}}, \bibinfo {author} {\bibfnamefont {J.~M.}\ \bibnamefont
  {Lattimer}}, \bibinfo {author} {\bibfnamefont {C.~J.}\ \bibnamefont
  {Pethick}},\ and\ \bibinfo {author} {\bibfnamefont {A.}~\bibnamefont
  {Schwenk}},\ }\bibfield  {title} {\bibinfo {title} {{Equation of state and
  neutron star properties constrained by nuclear physics and observation}},\
  }\href {https://doi.org/10.1088/0004-637X/773/1/11} {\bibfield  {journal}
  {\bibinfo  {journal} {Astrophys. J.}\ }\textbf {\bibinfo {volume} {773}},\
  \bibinfo {pages} {11} (\bibinfo {year} {2013})}\BibitemShut {NoStop}%
\bibitem [{\citenamefont {Annala}\ \emph {et~al.}(2018)\citenamefont {Annala},
  \citenamefont {Gorda}, \citenamefont {Kurkela},\ and\ \citenamefont
  {Vuorinen}}]{Anna18TidalDef}%
  \BibitemOpen
  \bibfield  {author} {\bibinfo {author} {\bibfnamefont {E.}~\bibnamefont
  {Annala}}, \bibinfo {author} {\bibfnamefont {T.}~\bibnamefont {Gorda}},
  \bibinfo {author} {\bibfnamefont {A.}~\bibnamefont {Kurkela}},\ and\ \bibinfo
  {author} {\bibfnamefont {A.}~\bibnamefont {Vuorinen}},\ }\bibfield  {title}
  {\bibinfo {title} {Gravitational-wave constraints on the neutron-star-matter
  equation of state},\ }\href {https://doi.org/10.1103/PhysRevLett.120.172703}
  {\bibfield  {journal} {\bibinfo  {journal} {Phys. Rev. Lett.}\ }\textbf
  {\bibinfo {volume} {120}},\ \bibinfo {pages} {172703} (\bibinfo {year}
  {2018})}\BibitemShut {NoStop}%
\bibitem [{\citenamefont {Most}\ \emph {et~al.}(2018)\citenamefont {Most},
  \citenamefont {Weih}, \citenamefont {Rezzolla},\ and\ \citenamefont
  {Schaffner-Bielich}}]{Most18TidalDef}%
  \BibitemOpen
  \bibfield  {author} {\bibinfo {author} {\bibfnamefont {E.~R.}\ \bibnamefont
  {Most}}, \bibinfo {author} {\bibfnamefont {L.~R.}\ \bibnamefont {Weih}},
  \bibinfo {author} {\bibfnamefont {L.}~\bibnamefont {Rezzolla}},\ and\
  \bibinfo {author} {\bibfnamefont {J.}~\bibnamefont {Schaffner-Bielich}},\
  }\bibfield  {title} {\bibinfo {title} {{New constraints on radii and tidal
  deformabilities of neutron stars from GW170817}},\ }\href
  {https://doi.org/10.1103/PhysRevLett.120.261103} {\bibfield  {journal}
  {\bibinfo  {journal} {Phys. Rev. Lett.}\ }\textbf {\bibinfo {volume} {120}},\
  \bibinfo {pages} {261103} (\bibinfo {year} {2018})}\BibitemShut {NoStop}%
\bibitem [{\citenamefont {Tews}\ \emph {et~al.}(2018)\citenamefont {Tews},
  \citenamefont {Margueron},\ and\ \citenamefont {Reddy}}]{Tews18TidalDef}%
  \BibitemOpen
  \bibfield  {author} {\bibinfo {author} {\bibfnamefont {I.}~\bibnamefont
  {Tews}}, \bibinfo {author} {\bibfnamefont {J.}~\bibnamefont {Margueron}},\
  and\ \bibinfo {author} {\bibfnamefont {S.}~\bibnamefont {Reddy}},\ }\bibfield
   {title} {\bibinfo {title} {{Critical examination of constraints on the
  equation of state of dense matter obtained from GW170817}},\ }\href
  {https://doi.org/10.1103/PhysRevC.98.045804} {\bibfield  {journal} {\bibinfo
  {journal} {Phys. Rev. C}\ }\textbf {\bibinfo {volume} {98}},\ \bibinfo
  {pages} {045804} (\bibinfo {year} {2018})}\BibitemShut {NoStop}%
\bibitem [{\citenamefont {Lim}\ and\ \citenamefont {Holt}(2018)}]{Lim2018}%
  \BibitemOpen
  \bibfield  {author} {\bibinfo {author} {\bibfnamefont {Y.}~\bibnamefont
  {Lim}}\ and\ \bibinfo {author} {\bibfnamefont {J.~W.}\ \bibnamefont {Holt}},\
  }\bibfield  {title} {\bibinfo {title} {{Neutron star tidal deformabilities
  constrained by nuclear theory and experiment}},\ }\href
  {https://doi.org/10.1103/PhysRevLett.121.062701} {\bibfield  {journal}
  {\bibinfo  {journal} {Phys. Rev. Lett.}\ }\textbf {\bibinfo {volume} {121}},\
  \bibinfo {pages} {062701} (\bibinfo {year} {2018})}\BibitemShut {NoStop}%
\bibitem [{\citenamefont {Capano}\ \emph {et~al.}(2020)\citenamefont {Capano},
  \citenamefont {Tews}, \citenamefont {Brown}, \citenamefont {Margalit},
  \citenamefont {De}, \citenamefont {Kumar}, \citenamefont {Brown},
  \citenamefont {Krishnan},\ and\ \citenamefont {Reddy}}]{Capa20NatAst}%
  \BibitemOpen
  \bibfield  {author} {\bibinfo {author} {\bibfnamefont {C.~D.}\ \bibnamefont
  {Capano}}, \bibinfo {author} {\bibfnamefont {I.}~\bibnamefont {Tews}},
  \bibinfo {author} {\bibfnamefont {S.~M.}\ \bibnamefont {Brown}}, \bibinfo
  {author} {\bibfnamefont {B.}~\bibnamefont {Margalit}}, \bibinfo {author}
  {\bibfnamefont {S.}~\bibnamefont {De}}, \bibinfo {author} {\bibfnamefont
  {S.}~\bibnamefont {Kumar}}, \bibinfo {author} {\bibfnamefont {D.~A.}\
  \bibnamefont {Brown}}, \bibinfo {author} {\bibfnamefont {B.}~\bibnamefont
  {Krishnan}},\ and\ \bibinfo {author} {\bibfnamefont {S.}~\bibnamefont
  {Reddy}},\ }\bibfield  {title} {\bibinfo {title} {{Stringent constraints on
  neutron-star radii from multimessenger observations and nuclear theory}},\
  }\href {https://doi.org/10.1038/s41550-020-1014-6} {\bibfield  {journal}
  {\bibinfo  {journal} {Nature Astron.}\ }\textbf {\bibinfo {volume} {4}},\
  \bibinfo {pages} {625} (\bibinfo {year} {2020})}\BibitemShut {NoStop}%
\bibitem [{\citenamefont {Essick}\ \emph {et~al.}(2020)\citenamefont {Essick},
  \citenamefont {Tews}, \citenamefont {Landry}, \citenamefont {Reddy},\ and\
  \citenamefont {Holz}}]{Essick2020}%
  \BibitemOpen
  \bibfield  {author} {\bibinfo {author} {\bibfnamefont {R.}~\bibnamefont
  {Essick}}, \bibinfo {author} {\bibfnamefont {I.}~\bibnamefont {Tews}},
  \bibinfo {author} {\bibfnamefont {P.}~\bibnamefont {Landry}}, \bibinfo
  {author} {\bibfnamefont {S.}~\bibnamefont {Reddy}},\ and\ \bibinfo {author}
  {\bibfnamefont {D.~E.}\ \bibnamefont {Holz}},\ }\bibfield  {title} {\bibinfo
  {title} {{Direct Astrophysical Tests of Chiral Effective Field Theory at
  Supranuclear Densities}},\ }\href
  {https://doi.org/10.1103/PhysRevC.102.055803} {\bibfield  {journal} {\bibinfo
   {journal} {Phys. Rev. C}\ }\textbf {\bibinfo {volume} {102}},\ \bibinfo
  {pages} {055803} (\bibinfo {year} {2020})}\BibitemShut {NoStop}%
\bibitem [{\citenamefont {Dietrich}\ \emph {et~al.}(2020)\citenamefont
  {Dietrich}, \citenamefont {Coughlin}, \citenamefont {Pang}, \citenamefont
  {Bulla}, \citenamefont {Heinzel}, \citenamefont {Issa}, \citenamefont
  {Tews},\ and\ \citenamefont {Antier}}]{Dietrich2020}%
  \BibitemOpen
  \bibfield  {author} {\bibinfo {author} {\bibfnamefont {T.}~\bibnamefont
  {Dietrich}}, \bibinfo {author} {\bibfnamefont {M.~W.}\ \bibnamefont
  {Coughlin}}, \bibinfo {author} {\bibfnamefont {P.~T.~H.}\ \bibnamefont
  {Pang}}, \bibinfo {author} {\bibfnamefont {M.}~\bibnamefont {Bulla}},
  \bibinfo {author} {\bibfnamefont {J.}~\bibnamefont {Heinzel}}, \bibinfo
  {author} {\bibfnamefont {L.}~\bibnamefont {Issa}}, \bibinfo {author}
  {\bibfnamefont {I.}~\bibnamefont {Tews}},\ and\ \bibinfo {author}
  {\bibfnamefont {S.}~\bibnamefont {Antier}},\ }\bibfield  {title} {\bibinfo
  {title} {{Multimessenger constraints on the neutron-star equation of state
  and the Hubble constant}},\ }\href {https://doi.org/10.1126/science.abb4317}
  {\bibfield  {journal} {\bibinfo  {journal} {Science}\ }\textbf {\bibinfo
  {volume} {370}},\ \bibinfo {pages} {1450} (\bibinfo {year}
  {2020})}\BibitemShut {NoStop}%
\bibitem [{\citenamefont {Raaijmakers}\ \emph {et~al.}(2021)\citenamefont
  {Raaijmakers}, \citenamefont {Greif}, \citenamefont {Hebeler}, \citenamefont
  {Hinderer}, \citenamefont {Nissanke}, \citenamefont {Schwenk}, \citenamefont
  {Riley}, \citenamefont {Watts}, \citenamefont {Lattimer},\ and\ \citenamefont
  {Ho}}]{Raaijmakers2021}%
  \BibitemOpen
  \bibfield  {author} {\bibinfo {author} {\bibfnamefont {G.}~\bibnamefont
  {Raaijmakers}}, \bibinfo {author} {\bibfnamefont {S.~K.}\ \bibnamefont
  {Greif}}, \bibinfo {author} {\bibfnamefont {K.}~\bibnamefont {Hebeler}},
  \bibinfo {author} {\bibfnamefont {T.}~\bibnamefont {Hinderer}}, \bibinfo
  {author} {\bibfnamefont {S.}~\bibnamefont {Nissanke}}, \bibinfo {author}
  {\bibfnamefont {A.}~\bibnamefont {Schwenk}}, \bibinfo {author} {\bibfnamefont
  {T.~E.}\ \bibnamefont {Riley}}, \bibinfo {author} {\bibfnamefont {A.~L.}\
  \bibnamefont {Watts}}, \bibinfo {author} {\bibfnamefont {J.~M.}\ \bibnamefont
  {Lattimer}},\ and\ \bibinfo {author} {\bibfnamefont {W.~C.~G.}\ \bibnamefont
  {Ho}},\ }\bibfield  {title} {\bibinfo {title} {{Constraints on the Dense
  Matter Equation of State and Neutron Star Properties from
  NICER\textquoteright{}s Mass\textendash{}Radius Estimate of PSR J0740+6620
  and Multimessenger Observations}},\ }\href
  {https://doi.org/10.3847/2041-8213/ac089a} {\bibfield  {journal} {\bibinfo
  {journal} {Astrophys. J. Lett.}\ }\textbf {\bibinfo {volume} {918}},\
  \bibinfo {pages} {L29} (\bibinfo {year} {2021})}\BibitemShut {NoStop}%
\bibitem [{\citenamefont {Essick}\ \emph {et~al.}(2021)\citenamefont {Essick},
  \citenamefont {Tews}, \citenamefont {Landry},\ and\ \citenamefont
  {Schwenk}}]{Essick2021}%
  \BibitemOpen
  \bibfield  {author} {\bibinfo {author} {\bibfnamefont {R.}~\bibnamefont
  {Essick}}, \bibinfo {author} {\bibfnamefont {I.}~\bibnamefont {Tews}},
  \bibinfo {author} {\bibfnamefont {P.}~\bibnamefont {Landry}},\ and\ \bibinfo
  {author} {\bibfnamefont {A.}~\bibnamefont {Schwenk}},\ }\bibfield  {title}
  {\bibinfo {title} {{Astrophysical Constraints on the Symmetry Energy and the
  Neutron Skin of $^{208}$Pb with Minimal Modeling Assumptions}},\ }\href
  {https://doi.org/10.1103/PhysRevLett.127.192701} {\bibfield  {journal}
  {\bibinfo  {journal} {Phys. Rev. Lett.}\ }\textbf {\bibinfo {volume} {127}},\
  \bibinfo {pages} {192701} (\bibinfo {year} {2021})}\BibitemShut {NoStop}%
\bibitem [{\citenamefont {Drischler}\ \emph
  {et~al.}(2021{\natexlab{b}})\citenamefont {Drischler}, \citenamefont {Han},
  \citenamefont {Lattimer}, \citenamefont {Prakash}, \citenamefont {Reddy},\
  and\ \citenamefont {Zhao}}]{DrischlerHan2021}%
  \BibitemOpen
  \bibfield  {author} {\bibinfo {author} {\bibfnamefont {C.}~\bibnamefont
  {Drischler}}, \bibinfo {author} {\bibfnamefont {S.}~\bibnamefont {Han}},
  \bibinfo {author} {\bibfnamefont {J.~M.}\ \bibnamefont {Lattimer}}, \bibinfo
  {author} {\bibfnamefont {M.}~\bibnamefont {Prakash}}, \bibinfo {author}
  {\bibfnamefont {S.}~\bibnamefont {Reddy}},\ and\ \bibinfo {author}
  {\bibfnamefont {T.}~\bibnamefont {Zhao}},\ }\bibfield  {title} {\bibinfo
  {title} {{Limiting masses and radii of neutron stars and their
  implications}},\ }\href {https://doi.org/10.1103/PhysRevC.103.045808}
  {\bibfield  {journal} {\bibinfo  {journal} {Phys. Rev. C}\ }\textbf {\bibinfo
  {volume} {103}},\ \bibinfo {pages} {045808} (\bibinfo {year}
  {2021}{\natexlab{b}})}\BibitemShut {NoStop}%
\bibitem [{\citenamefont {Huth}\ \emph {et~al.}(2022)\citenamefont {Huth},
  \citenamefont {Pang}, \citenamefont {Tews}, \citenamefont {Dietrich},
  \citenamefont {Le~F\`evre}, \citenamefont {Schwenk}, \citenamefont
  {Trautmann}, \citenamefont {Agarwal}, \citenamefont {Bulla}, \citenamefont
  {Coughlin},\ and\ \citenamefont {Van Den~Broeck}}]{Huth2021}%
  \BibitemOpen
  \bibfield  {author} {\bibinfo {author} {\bibfnamefont {S.}~\bibnamefont
  {Huth}}, \bibinfo {author} {\bibfnamefont {P.~T.~H.}\ \bibnamefont {Pang}},
  \bibinfo {author} {\bibfnamefont {I.}~\bibnamefont {Tews}}, \bibinfo {author}
  {\bibfnamefont {T.}~\bibnamefont {Dietrich}}, \bibinfo {author}
  {\bibfnamefont {A.}~\bibnamefont {Le~F\`evre}}, \bibinfo {author}
  {\bibfnamefont {A.}~\bibnamefont {Schwenk}}, \bibinfo {author} {\bibfnamefont
  {W.}~\bibnamefont {Trautmann}}, \bibinfo {author} {\bibfnamefont
  {K.}~\bibnamefont {Agarwal}}, \bibinfo {author} {\bibfnamefont
  {M.}~\bibnamefont {Bulla}}, \bibinfo {author} {\bibfnamefont {M.~W.}\
  \bibnamefont {Coughlin}},\ and\ \bibinfo {author} {\bibfnamefont
  {C.}~\bibnamefont {Van Den~Broeck}},\ }\bibfield  {title} {\bibinfo {title}
  {{Constraining Neutron-Star Matter with Microscopic and Macroscopic
  Collisions}},\ }\href {https://doi.org/10.1038/s41586-022-04750-w} {\bibfield
   {journal} {\bibinfo  {journal} {Nature}\ }\textbf {\bibinfo {volume}
  {606}},\ \bibinfo {pages} {276} (\bibinfo {year} {2022})}\BibitemShut
  {NoStop}%
\bibitem [{\citenamefont {Horowitz}\ and\ \citenamefont
  {Schwenk}(2006{\natexlab{a}})}]{HorowitzSchwenk2006a}%
  \BibitemOpen
  \bibfield  {author} {\bibinfo {author} {\bibfnamefont {C.~J.}\ \bibnamefont
  {Horowitz}}\ and\ \bibinfo {author} {\bibfnamefont {A.}~\bibnamefont
  {Schwenk}},\ }\bibfield  {title} {\bibinfo {title} {The virial equation of
  state of low-density neutron matter},\ }\href
  {https://doi.org/https://doi.org/10.1016/j.physletb.2006.05.055} {\bibfield
  {journal} {\bibinfo  {journal} {Phys. Lett. B}\ }\textbf {\bibinfo {volume}
  {638}},\ \bibinfo {pages} {153} (\bibinfo {year}
  {2006}{\natexlab{a}})}\BibitemShut {NoStop}%
\bibitem [{\citenamefont {Horowitz}\ and\ \citenamefont
  {Schwenk}(2006{\natexlab{b}})}]{HorowitzSchwenk2006b}%
  \BibitemOpen
  \bibfield  {author} {\bibinfo {author} {\bibfnamefont {C.~J.}\ \bibnamefont
  {Horowitz}}\ and\ \bibinfo {author} {\bibfnamefont {A.}~\bibnamefont
  {Schwenk}},\ }\bibfield  {title} {\bibinfo {title} {Cluster formation and the
  virial equation of state of low-density nuclear matter},\ }\href
  {https://doi.org/https://doi.org/10.1016/j.nuclphysa.2006.05.009} {\bibfield
  {journal} {\bibinfo  {journal} {Nucl. Phys. A}\ }\textbf {\bibinfo {volume}
  {776}},\ \bibinfo {pages} {55} (\bibinfo {year}
  {2006}{\natexlab{b}})}\BibitemShut {NoStop}%
\bibitem [{\citenamefont {Fetter}\ and\ \citenamefont
  {Walecka}(1972)}]{FetterWalecka}%
  \BibitemOpen
  \bibfield  {author} {\bibinfo {author} {\bibfnamefont {A.~L.}\ \bibnamefont
  {Fetter}}\ and\ \bibinfo {author} {\bibfnamefont {J.~D.}\ \bibnamefont
  {Walecka}},\ }\href@noop {} {\emph {\bibinfo {title} {{Quantum Theory of
  Many-Particle Systems}}}}\ (\bibinfo  {publisher} {McGraw-Hill, New York},\
  \bibinfo {year} {1972})\BibitemShut {NoStop}%
\bibitem [{\citenamefont {Negele}\ and\ \citenamefont
  {Orland}(1998)}]{NegeleOrland}%
  \BibitemOpen
  \bibfield  {author} {\bibinfo {author} {\bibfnamefont {J.~W.}\ \bibnamefont
  {Negele}}\ and\ \bibinfo {author} {\bibfnamefont {H.}~\bibnamefont
  {Orland}},\ }\href@noop {} {\emph {\bibinfo {title} {{Quantum Many-Particle
  Systems}}}}\ (\bibinfo  {publisher} {Westview Press, Boulder},\ \bibinfo
  {year} {1998})\BibitemShut {NoStop}%
\bibitem [{\citenamefont {Entem}\ \emph {et~al.}(2017)\citenamefont {Entem},
  \citenamefont {Machleidt},\ and\ \citenamefont {Nosyk}}]{EMN2017}%
  \BibitemOpen
  \bibfield  {author} {\bibinfo {author} {\bibfnamefont {D.~R.}\ \bibnamefont
  {Entem}}, \bibinfo {author} {\bibfnamefont {R.}~\bibnamefont {Machleidt}},\
  and\ \bibinfo {author} {\bibfnamefont {Y.}~\bibnamefont {Nosyk}},\ }\bibfield
   {title} {\bibinfo {title} {High-quality two-nucleon potentials up to fifth
  order of the chiral expansion},\ }\href
  {https://doi.org/10.1103/PhysRevC.96.024004} {\bibfield  {journal} {\bibinfo
  {journal} {Phys. Rev. C}\ }\textbf {\bibinfo {volume} {96}},\ \bibinfo
  {pages} {024004} (\bibinfo {year} {2017})}\BibitemShut {NoStop}%
\bibitem [{\citenamefont {Hebeler}(2021)}]{Hebeler2021}%
  \BibitemOpen
  \bibfield  {author} {\bibinfo {author} {\bibfnamefont {K.}~\bibnamefont
  {Hebeler}},\ }\bibfield  {title} {\bibinfo {title} {{Three-nucleon forces:
  Implementation and applications to atomic nuclei and dense matter}},\ }\href
  {https://doi.org/10.1016/j.physrep.2020.08.009} {\bibfield  {journal}
  {\bibinfo  {journal} {Phys. Rept.}\ }\textbf {\bibinfo {volume} {890}},\
  \bibinfo {pages} {1} (\bibinfo {year} {2021})}\BibitemShut {NoStop}%
\bibitem [{\citenamefont {Kohn}\ and\ \citenamefont
  {Luttinger}(1960)}]{KohnLuttinger1960}%
  \BibitemOpen
  \bibfield  {author} {\bibinfo {author} {\bibfnamefont {W.}~\bibnamefont
  {Kohn}}\ and\ \bibinfo {author} {\bibfnamefont {J.~M.}\ \bibnamefont
  {Luttinger}},\ }\bibfield  {title} {\bibinfo {title} {Ground-state energy of
  a many-fermion system},\ }\href {https://doi.org/10.1103/PhysRev.118.41}
  {\bibfield  {journal} {\bibinfo  {journal} {Phys. Rev.}\ }\textbf {\bibinfo
  {volume} {118}},\ \bibinfo {pages} {41} (\bibinfo {year} {1960})}\BibitemShut
  {NoStop}%
\bibitem [{\citenamefont {Wellenhofer}(2019)}]{Wellenhofer:2018qqw}%
  \BibitemOpen
  \bibfield  {author} {\bibinfo {author} {\bibfnamefont {C.}~\bibnamefont
  {Wellenhofer}},\ }\bibfield  {title} {\bibinfo {title} {{Zero-temperature
  limit and statistical quasiparticles in many-body perturbation theory}},\
  }\href {https://doi.org/10.1103/PhysRevC.99.065811} {\bibfield  {journal}
  {\bibinfo  {journal} {Phys. Rev. C}\ }\textbf {\bibinfo {volume} {99}},\
  \bibinfo {pages} {065811} (\bibinfo {year} {2019})}\BibitemShut {NoStop}%
\bibitem [{\citenamefont {Rasmussen}\ and\ \citenamefont
  {Williams}(2005)}]{Rasmussen2005}%
  \BibitemOpen
  \bibfield  {author} {\bibinfo {author} {\bibfnamefont {C.~E.}\ \bibnamefont
  {Rasmussen}}\ and\ \bibinfo {author} {\bibfnamefont {C.~K.~I.}\ \bibnamefont
  {Williams}},\ }\href@noop {} {\emph {\bibinfo {title} {Gaussian Processes for
  Machine Learning}}}\ (\bibinfo  {publisher} {MIT Press},\ \bibinfo {address}
  {Cambridge},\ \bibinfo {year} {2005})\BibitemShut {NoStop}%
\bibitem [{\citenamefont {Chilenski}\ \emph {et~al.}(2015)\citenamefont
  {Chilenski}, \citenamefont {Greenwald}, \citenamefont {Marzouk},
  \citenamefont {Howard}, \citenamefont {White}, \citenamefont {Rice},\ and\
  \citenamefont {Walk}}]{Chilenski_2015}%
  \BibitemOpen
  \bibfield  {author} {\bibinfo {author} {\bibfnamefont {M.~A.}\ \bibnamefont
  {Chilenski}}, \bibinfo {author} {\bibfnamefont {M.}~\bibnamefont
  {Greenwald}}, \bibinfo {author} {\bibfnamefont {Y.}~\bibnamefont {Marzouk}},
  \bibinfo {author} {\bibfnamefont {N.~T.}\ \bibnamefont {Howard}}, \bibinfo
  {author} {\bibfnamefont {A.~E.}\ \bibnamefont {White}}, \bibinfo {author}
  {\bibfnamefont {J.~E.}\ \bibnamefont {Rice}},\ and\ \bibinfo {author}
  {\bibfnamefont {J.~R.}\ \bibnamefont {Walk}},\ }\bibfield  {title} {\bibinfo
  {title} {Improved profile fitting and quantification of uncertainty in
  experimental measurements of impurity transport coefficients using gaussian
  process regression},\ }\href {https://doi.org/10.1088/0029-5515/55/2/023012}
  {\bibfield  {journal} {\bibinfo  {journal} {Nucl. Fusion}\ }\textbf {\bibinfo
  {volume} {55}},\ \bibinfo {pages} {023012} (\bibinfo {year}
  {2015})}\BibitemShut {NoStop}%
\bibitem [{\citenamefont {Epelbaum}\ \emph {et~al.}(2015)\citenamefont
  {Epelbaum}, \citenamefont {Krebs},\ and\ \citenamefont
  {Mei{\ss}ner}}]{EKM2015}%
  \BibitemOpen
  \bibfield  {author} {\bibinfo {author} {\bibfnamefont {E.}~\bibnamefont
  {Epelbaum}}, \bibinfo {author} {\bibfnamefont {H.}~\bibnamefont {Krebs}},\
  and\ \bibinfo {author} {\bibfnamefont {U.-G.}\ \bibnamefont {Mei{\ss}ner}},\
  }\bibfield  {title} {\bibinfo {title} {Improved chiral nucleon-nucleon
  potential up to next-to-next-to-next-to-leading order},\ }\href
  {https://doi.org/10.1140/epja/i2015-15053-8} {\bibfield  {journal} {\bibinfo
  {journal} {Eur. Phys. J. A}\ }\textbf {\bibinfo {volume} {51}},\ \bibinfo
  {pages} {53} (\bibinfo {year} {2015})}\BibitemShut {NoStop}%
\bibitem [{\citenamefont {Lattimer}\ and\ \citenamefont
  {Swesty}(1991)}]{Lattimer_Swesty_1991}%
  \BibitemOpen
  \bibfield  {author} {\bibinfo {author} {\bibfnamefont {J.~M.}\ \bibnamefont
  {Lattimer}}\ and\ \bibinfo {author} {\bibfnamefont {F.~D.}\ \bibnamefont
  {Swesty}},\ }\bibfield  {title} {\bibinfo {title} {{A Generalized equation of
  state for hot, dense matter}},\ }\href
  {https://doi.org/10.1016/0375-9474(91)90452-C} {\bibfield  {journal}
  {\bibinfo  {journal} {Nucl. Phys. A}\ }\textbf {\bibinfo {volume} {535}},\
  \bibinfo {pages} {331} (\bibinfo {year} {1991})}\BibitemShut {NoStop}%
\bibitem [{\citenamefont {Drischler}\ \emph
  {et~al.}(2020{\natexlab{b}})\citenamefont {Drischler}, \citenamefont
  {Melendez}, \citenamefont {Furnstahl},\ and\ \citenamefont
  {Phillips}}]{Drischler2020}%
  \BibitemOpen
  \bibfield  {author} {\bibinfo {author} {\bibfnamefont {C.}~\bibnamefont
  {Drischler}}, \bibinfo {author} {\bibfnamefont {J.~A.}\ \bibnamefont
  {Melendez}}, \bibinfo {author} {\bibfnamefont {R.~J.}\ \bibnamefont
  {Furnstahl}},\ and\ \bibinfo {author} {\bibfnamefont {D.~R.}\ \bibnamefont
  {Phillips}},\ }\bibfield  {title} {\bibinfo {title} {{Quantifying
  uncertainties and correlations in the nuclear-matter equation of state}},\
  }\href {https://doi.org/10.1103/PhysRevC.102.054315} {\bibfield  {journal}
  {\bibinfo  {journal} {Phys. Rev. C}\ }\textbf {\bibinfo {volume} {102}},\
  \bibinfo {pages} {054315} (\bibinfo {year} {2020}{\natexlab{b}})}\BibitemShut
  {NoStop}%
\end{thebibliography}%

\end{document}